\documentclass[12pt,article,aps, amsfonts, nofootinbib]{revtex4-2}

\usepackage{graphics,setspace,epsfig,color}
\usepackage[letterpaper, dvips,width=7.5in,height=8.5in,includemp=false]{geometry}

\usepackage{amsmath}
\usepackage{amsfonts}
\usepackage{amssymb}
\usepackage{amscd}
\usepackage{braket}
\usepackage{bm}
\usepackage{multirow}
\usepackage{booktabs}
\usepackage{caption}
\usepackage{subcaption}
\usepackage{physics}
\usepackage{stackrel}
\usepackage{tcolorbox}
\usepackage{epsf}
\usepackage{amsmath}
\usepackage{amsfonts}
\usepackage[utf8]{inputenc}
\usepackage{microtype}
\usepackage{amssymb}
\usepackage{braket} 
\usepackage{bbm}
\usepackage{bm} 
\usepackage{mathtools}
\usepackage{slashed}

\definecolor{myblue}  {rgb}{0.207031, 0.28125, 0.378906}
\definecolor{myred}  {rgb}{0.367188, 0.09375, 0.0742188}
\definecolor{mybeige}  {rgb}{0.96875, 0.957031, 0.933594}
\definecolor{mygreen}  {rgb}{0.746094, 0.773438, 0.28125}
\definecolor{mysepia}  {rgb}{0.421875, 0.257813, 0.109375}

\newcommand{\beq}{\begin{equation}}
\newcommand{\eeq}{\end{equation}}
\newcommand{\bea}{\begin{eqnarray}}
\newcommand{\eea}{\end{eqnarray}}

\newcommand{\eq}[1]{Eq.~\ref{#1}}
\newcommand{\fig}[1]{Fig.~\ref{#1}}
\newcommand{\tab}[1]{Table~\ref{#1}}
\newcommand{\sect}[1]{section~\ref{#1}}

\newcommand{\calE}{\mathcal{E}}

\newcommand{\calO}{\mathcal{O}}

\usepackage{amsmath, amssymb, amsthm}
\usepackage{mathtools}           
\usepackage{physics}             
\usepackage{slashed}             
\usepackage{bbm}                 
\usepackage{eucal}               
\usepackage{float}        

\usepackage{quantikz}            

\usepackage{graphicx}
\usepackage{booktabs}            
\usepackage{siunitx}             
\usepackage{geometry}

\usepackage[colorlinks=true, citecolor=blue, linkcolor=blue, urlcolor=blue]{hyperref}

\newcommand{\Om}{|\Omega\rangle}
\newcommand{\OmDagger}{\langle\Omega|}
\begin{document}

\title{Hamiltonian spectra in quantum computers through the generalized eigenvalue method} 
\author{Valery Simonyan}
\email{vsimon24@umd.edu}
\author{Paulo F. Bedaque}
\email{bedaque@umd.edu}
\author{ Gregory Ridgway}
\email{gridgwa1@umd.edu}
\affiliation{University of Maryland, College Park, MD, USA} 

\preprint{}
\begin{abstract}
Quantum computers can generate real-time correlators of field theories. By adapting the generalized eigenvalue problem to these correlators, energy eigenvalues can be extracted directly. The method is tested using both classical simulations and quantum hardware, successfully resolving several low-lying energy levels in agreement with exact diagonalization. Comparison with an alternative spectrum determination based on the Fourier transform of correlators shows that the proposed approach is substantially more efficient.
\end{abstract}
\maketitle

\newcommand{\R}{\mathbb{R}}
\newcommand{\one}{\mathbbm{1}}
\newcommand{\lag}{\mathcal{L}}
\newcommand{\ham}{\mathcal{H}}
\newcommand{\ord}[1]{\mathcal{O}\!\left(#1\right)}
\newcommand{\diag}{\operatorname{diag}}

\newcommand{\lam}[2]{\lambda_{#1}(#2)}

\section{Introduction}
\label{sec:intro}

The spectrum of a Hamiltonian is among the most basic pieces of information one
can ask about a quantum system. In nuclear and particle physics the energies of
the eigenstates are the masses of the particles, the binding energies of bound
states and, through the finite volume dependence of the levels, the scattering
amplitudes of the theory. Computing the low lying spectrum of a strongly coupled
field theory from first principles is, consequently, one of the central goals of
the field.

The standard non-perturbative tool for this task is the Monte Carlo evaluation of
Euclidean correlators. A correlator of interpolating operators behaves, at large
imaginary time, as a sum of decaying exponentials $\sum_n |\langle 0|O|n\rangle|^2
e^{-E_n\tau}$. This decay is a mixed blessing. The contribution of the excited
states dies away relative to that of the lowest state, so the ground state energy
can be extracted from a plateau at large $\tau$ with little knowledge of the
operator used. The same suppression, however, is what makes the excited states
difficult: their contribution has to be extracted from the small deviations of the
correlator from its asymptotic form, precisely in the region where the
signal-to-noise ratio of the stochastic estimate degrades exponentially.

A well established way of dealing with this problem is to consider not one
correlator but the matrix of correlators of a set of interpolating operators and
to solve a generalized eigenvalue problem (GEVP) for it \cite{LuscherWolff,michael,
blossier}. The eigenvectors are the linear combinations of the operators that
couple, to a good approximation, to a single eigenstate, and the corresponding
eigenvalues isolate the individual energies. The method is now routine in
lattice field theory, where it is the standard route to excited states and to the
extraction of phase shifts.

Quantum computers change the nature of the problem because they naturally produce
correlators in \emph{real} time. Real time correlators do not decay: all the states
created by the operator contribute with an undamped, oscillating weight. There is
no ground state dominance to rely on, but also no exponential loss of information
about the excited levels. The obvious way of reading the frequencies off such a
signal is to Fourier transform it, but the resolution of a transform is set by the
inverse of the total time extent, and resolving nearly degenerate levels --- or
even distinguishing a genuine peak from a ringing artifact --- requires evolution
times far beyond what current devices, with their finite coherence and accumulating
gate and Trotter errors, can deliver.

In this paper we adapt the generalized eigenvalue problem to real time correlators.
We describe how the correlator matrix of a set of Pauli operators can be measured
with a single ancilla qubit, and test the resulting procedure on the fuzzy
$\sigma$-model \cite{fuzzy}, using exact time evolution, noisy simulations of
realistic devices, and IonQ's Forte-Enterprise quantum computer. A previous attempt to extract the gap of this model using quantum hardware simulations was made in \cite{fuzzy-freedom} and motivated the present work. We find that, with the present methods,
several low lying levels can be resolved in agreement with exact diagonalization
at time extents where the Fourier transform of the same data is useless.

\section{Generalized Eigenvalue Problem}
\label{sec:GEM}
Consider a system with Hamiltonian $\hat{H}$ and a discrete spectrum of $N$ eigenstates $|n\rangle$ with energies $E_n$.
We shift the Hamiltonian by an irrelevant constant so that the ground state $|0\rangle$ has energy $E_0=0$.
The energy eigenstates can be generated by acting on the ground state with the $N$ operators 
$\bar{O}_n = |n\rangle\langle 0| +  |0\rangle\langle n|$. These
\textit{perfect operators}
 have time correlators
\beq
\bra{0} \bar{O}_n(t) \bar{O}^\dagger_n(0)\ket{0} 
=
e^{-i E_n t}
\eeq 
that are single exponentials.
Of course, explicitly constructing these operators is equivalent to solving the model so they and the ground state are not, in general, available. 
What may be available is a smaller set of $M<N$ \textit{imperfect} operators $O_i$ and an approximate ground state $\ket{\Omega}$ that together generate approximate eigenstates:
\beq
    O_i \ket{\Omega} \approx \ket{i}, i=1,\cdots, M.
\label{eq:Oi}
\eeq 
In fact, it will be enough to have a set of imperfect operators that span the space of operators satisfying \eq{eq:Oi}.
Since the perfect operators can generate any eigenstate, one can write the imperfect operators as a linear combination of the perfect ones:
\beq
    O_i = \sum_n R_{in} \bar{O}_n 
\eeq 
where $R_{in}$ is an $M\times N$ matrix. We will discuss below practical ways to find good $O_i$ and $\ket{\Omega}$. Consider the matrix of correlators
\beq
    C_{ij}(t) = \OmDagger O_i(t) O_j^\dagger(0) \Om.
    \label{eq:Cij}
\eeq
$C_{ij}(t)$ can be expressed in terms of the perfect operators as
\begin{equation}\label{eq:Cij}
C_{ij}(t) \approx \sum_{m,n=1}^{N} R_{im} \OmDagger \bar{O}_m(t) \bar{O}^\dagger_n(0) \Om R^*_{jn} =  
    \sum_{m,n=1}^{N} R_{im} e^{-i E_n t} \delta_{mn} R^*_{jn} = (R U(t) R^\dagger)_{ij}
\end{equation}
where $U(t) = \diag(e^{-i E_1 t}, \ldots, e^{-i E_N t})$ is the time evolution matrix in the energy eigenbasis.  We now split the sum in \eq{eq:Cij} into the first $M$ states well approximated by the imperfect operators and the rest:

\begin{equation}
\begin{split}
    C_{ij}(t) &= \sum_{k} Q_{ik} e^{-i E_k t} Q^\dagger_{kj} + \sum_{a} P_{ia} e^{-i E_a t} P^\dagger_{aj}=(Q U_\text{in}(t) Q^\dagger)_{ij} + (P U_\text{out}(t) P^\dagger)_{ij} \\
              &= (Q U_\text{in}(t) Q^\dagger)_{ik}
              \Big[\mathbb{I} +
               \frac{1}{Q U_\text{in}(t) Q^\dagger} P U_\text{out}(t) P^\dagger\Big]_{kj}=(Q U_\text{in}(t) Q^\dagger)_{ij} + \calO\large((Q^{-1}P)^2\large),
\end{split}
\end{equation}
where $Q$ is an $M\times M$ matrix which we assume to be invertible, $P$ is an $M\times (N-M)$ matrix.
By our assumption, elements of $P$ are smaller than those of $Q$ so can be neglected. 
$U_\text{in}(t)$ and $U_\text{out}(t)$ are unitary diagonal matrices containing the time evolution factors of states with large and small overlap with the imperfect operator basis, respectively. 

We may now consider the inverse of the correlator matrix
\begin{equation}
\begin{split}
    C^{-1}(t) &= \Big[\mathbb{I}+\frac{1}{Q U_\text{in}(t) Q^\dagger} P U_\text{out}(t) P^\dagger\Big]^{-1} \frac{1}{Q U_\text{in}(t) Q^\dagger},
\end{split}
\end{equation}
which we can expand to find
\begin{equation}
\begin{split}
    C^{-1}(0) C(t) &= \frac{1}{Q^\dagger} U_\text{in}(t) Q^\dagger + \frac{1}{Q Q^\dagger} P U_\text{out}(t) P^\dagger - \frac{1}{Q Q^\dagger} P P^\dagger \frac{1}{Q^\dagger} U_\text{in}(t) Q^\dagger +...\\
    &= \frac{1}{Q^\dagger} U_\text{in}(t) Q^\dagger + \calO((Q^{-1}P)^2) \, ,
\end{split}
\end{equation}
where we have used $U_\text{in,out}(0) = \mathbb{I}$.
We may diagonalize the first term with eigenvectors $v^{(0)}_k$ satisfying $(Q^\dagger)^{-1} U_\text{in}(t) Q^\dagger v^{(0)}_k = \lambda^{(0)}_k(t) v^{(0)}_k$.
This is satisfied by $v_k = (Q^\dagger)^{-1} e_k$ where $U_\text{in}e_k = \lambda^{(0)}_k e_k$ and $\lambda^{(0)}_k(t)=e^{-i E_k t}$.
Using perturbation theory we can find the leading order correction to the eigenvalues as
\begin{equation}
    \lambda^{(1)}_k(t) = e^T_k Q^{-1} P (U_\text{out}(t) - \lambda^{(0)}_k(t) \mathbb{I}) P^\dagger (Q^\dagger)^{-1} e_k = \order{(Q^{-1}P)^2}
\end{equation}
which is small by construction. 
The eigenvalues of $C^{-1}(0) C(t)$ are then given by $\lambda_k(t) = \lambda^{(0)}_k(t) + \lambda^{(1)}_k(t) + ... = e^{-i E_k t} + \order{(Q^{-1}P)^2}$. Therefore,
 as long as linear combinations of the imperfect states $O_i\ket{\Omega}$ are close enough to the energy eigenstates $\ket{i}$ and include all the states contributing significantly to $C_{ij}$, the exact energy eigenvalues are approximately given by 

\begin{equation}\label{eq:log-energies}
    E_k^\text{eff}(t) = \frac{i}{t} \log(\lambda_k(t)) = E_k + \frac{i}{t}\log(1 + e^{i E_k t}\lambda^{(1)}_k(t)+...) = E_k + \frac{i}{t}\order{(Q^{-1}P)^2}
\end{equation}
where the correction term decays with time as $1/t$, or even faster due to destructive interference.

As a benchmark, we compare this generalized eigenvalue problem to an alternative method for extracting frequencies from an oscillatory signal, the Fourier transform. Consider the trace of the correlator matrix
\beq
    \Tr C(t)
    = \sum_i\OmDagger O_i(t) O_i^\dagger(0) \Om 
    = \sum_{i,n}\left|\OmDagger O_i(0) \ket{n} \right|^2 e^{-i E_n t} \, .
\eeq
To extract the energies, define the spectral density:
\begin{equation}
    S_\eta(\omega)
    = \int_{-\infty}^{\infty} dt \, e^{i \omega t - \eta |t|} \,\Tr C(t)
    = 2 \sum_{i,n}\left|\OmDagger O_i(0) \ket{n} \right|^2 \frac{\eta}{(\omega - E_n)^2 + \eta^2}
\label{eq:cont_ft}
\end{equation}
where we used $C_{ii}^*(t) = \OmDagger O_i(0) O_i^\dagger(t) \Om = C_{ii}(-t)$ for all diagonal elements, allowing us to use only the positive-time part of the signal that we actually measure. In reality, we only generate a finite set of time samples with a minimum time resolution $\Delta t$ and a maximum time extent $T$. We use the $\eta$ damping factor to handle Gibbs phenomena, i.e. ringing artifacts from our finitely sampled signal. We then approximate the above continuous expression via the trapezoid rule
\beq
    S^{\text{disc}}_\eta(\omega) = 2 \Delta t \, \text{Re}\, \text{Tr}\left(\frac{C(0) + C(T)e^{i \omega T - \eta T}}{2} + \sum_{k=1}^{N-1} e^{i \omega t_k - \eta t_k} C(t_k) \right)
\label{eq:disc_ft}
\eeq
In the limits $T\to \infty$ and $\Delta t \to 0$ we expect $S^{\text{disc}}_{\eta}(\omega) = S_{\eta}(\omega) + \ord{\Delta t^2}$, and to have well defined peaks at each energy level accessible to the $O_i$ operators.

\section{Calculating Off-Diagonal Correlators on a Quantum Computer}
\label{sec:circuitry}
It is straightforward to compute quantities of the form $\bra{\psi(t)}  A \ket{\psi(t)} = \bra{\psi(0)} U^\dagger(t) A U(t)\ket{\psi(0)}$ in a quantum computer. One prepares the state $\ket{\psi(0)}$, evolves it  by a time $t$ and computes the expectation value of $A$. But
it is not immediately obvious how to compute the elements of the correlator matrix $C_{ij}(t) = \bra{\Omega}  O_i(t) O^\dagger_j(0)\ket{\Omega}$. We will present now a method to compute these matrix elements in the case where the $O_i$ are Pauli matrices. As we will see, this is all that is required to use the generalized eigenvalue method.

The algorithm to compute $\bra{\Omega}  P_a(t) P_b(0)\ket{\Omega}$ is:
\begin{figure}[ht]
\centering
\begin{quantikz}
  \lstick{$\ket{0}$}   & \gate[2]{\calE} & \qw               & \meter{} &  \\
  \lstick{$\ket{\Omega}$}     &             & \gate{e^{-iHt}}   & \meter{} & \qw
\end{quantikz}
\end{figure}

\begin{itemize}
    \item Prepare the approximate ground state $\ket{\Omega}$ and an ancilla qubit in the state $\ket{0}$.
    \item Entangle the system with the ancilla using $\calE=e^{-i \frac{\pi}{4} X \otimes P_b}$, bringing the system state to 
          \begin{equation}
            \ket{\psi}=\frac{1}{\sqrt{2}}\ket{0}\otimes \ket{\Omega}-\frac{i}{\sqrt{2}}\ket{1}\otimes(P_b\ket{\Omega})
          \end{equation}        
    \item Time-evolve the system by applying $e^{-i (\mathbb{I} \otimes H )t}$, giving us the state
          \begin{equation}
            \ket{\psi(t)} = \frac{1}{\sqrt{2}}( \ket{0}\otimes e^{-i H t}\ket{\Omega})-\frac{i}{\sqrt{2}}\ket{1}\otimes (e^{-i H t}P_b\ket{\Omega})
          \end{equation}
    \item Bring the system into the diagonal basis to measure first the imaginary part of the correlator $\langle X \otimes P_a \rangle$ and then the real part $\langle Y \otimes P_a \rangle$.
    \item Finally, combine the two parts of the correlator to give 
    \begin{equation}
        \bra{\Omega}P_a(t) P_b(0) \ket{\Omega} = i\bra{\psi(t)} X \otimes P_a \ket{\psi(t)} - \bra{\psi(t)} Y \otimes P_a \ket{\psi(t)}
    \end{equation} 
\end{itemize}

This procedure has to be repeated for several pairs of Pauli matrices $P_a, P_b$. The linear combinations of operators that bring $\ket{\Omega}$ to states near the energy eigenstates are automatically generated by the diagonalization procedure in the generalized eigenvalue method. Of course, for the method to be useful in practice one needs a large enough set of Pauli matrices, large enough to be able to approximate the perfect ones. The symmetries of the problem and some qualitative insight into its dynamics reduces the number of necessary Pauli operators, as we will see in the example discussed in the next section.
Also, we may measure qubit-wise commuting Pauli operators simultaneously, thus reducing the number of circuits needed to prepare $|\psi(t)\rangle$.

\section{Generalized Eigenvalues in action: the Fuzzy $\sigma$-Model}
\label{sec:fuzzy_model}
In order to exemplify the method of the previous section we choose the
 fuzzy $\sigma$-model \cite{fuzzy, fuzzy-2,fuzzy-3,fuzzy-freedom}, which is equivalent to the Heisenberg comb \cite{comb-1,comb-2,comb-3,comb-5}. The construction of this model was motivated by the fact that it has the same continuum limit as the  $1+1$ dimensional $O(3) \  \sigma$-model and so provides a ``qubitization" or ``field space discretization" of the continuous theory.
 It shares several features with more complex field theories of  physical interest like asymptotic freedom and dynamical mass generation. 
 The model is defined on a one-dimensional lattice with $L$ sites and two qubits per site. These qubits are indexed by $x=0, \cdots 2L-1$ with the  even and odd values of $x$ referred to as the ``head" and the ``fuzz" qubits, respectively. The Hamiltonian of the model is:
\begin{equation}
    H = \sqrt\eta \left[\frac{g^2}{2}\sum_{x=0}^{L-1} \boldsymbol{\sigma}_{2x}\cdot \boldsymbol{\sigma}_{2x+1}+\frac{1}{3g^2}\sum_{x=0}^{L-1} \boldsymbol{\sigma}_{2x}\cdot \boldsymbol{\sigma}_{2x+2}\right]
\label{eq:H}
\end{equation}
where periodic boundary conditions $\sigma_{i+2L} = \sigma_i$ are implied. We are using units in which $\hbar$, the speed of light, and the lattice spacing are 1. 
The first term couples the head and fuzz qubits (kinetic term) at the same site while the second couples nearest-neighbor head qubits (potential term) \footnote{The motivation for the names ``kinetic" and ``potential" are discussed in \cite{fuzzy}.}, both with the anti-ferromagnetic sign favoring spin anti-alignment.
The parameter $\eta$ does not change the energy eigenstates and just scales the the eigenvalues; 
for simplicity we set $\eta=1$ \footnote{The value of $\sqrt\eta$ is adjusted so the spatial correlation lengths are the inverse of the energy gaps, ensuring Lorentz invariance. We will not be concerned with this step here.}.

\begin{figure}
    \centering
    \includegraphics[width=0.5\linewidth]{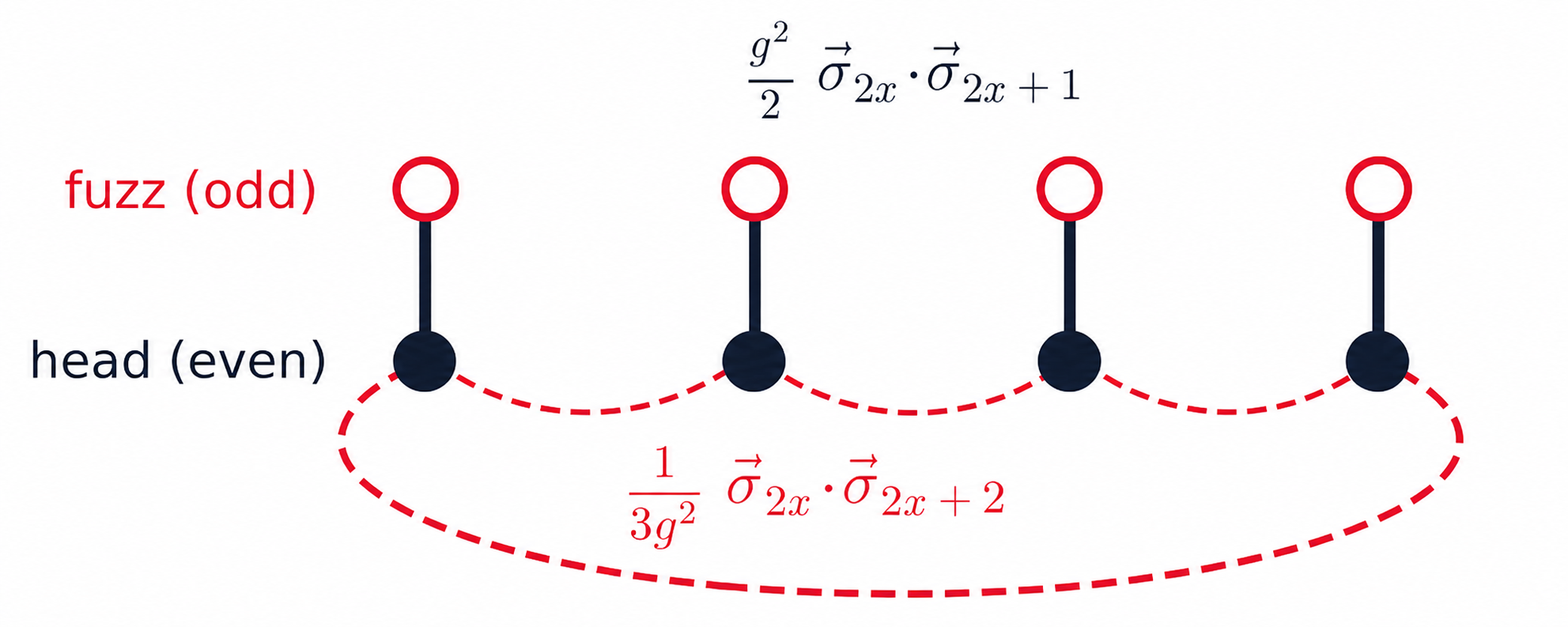}
    \caption{Visualization of the Hamiltonian in \eq{eq:H}.}
    \label{fig:placeholder}
\end{figure}

We can build some intuition about the model that will turn out to be useful when looking for a good approximation to the ground state.  Consider the strong coupling limit,  $g\to\infty$. In this limit, the Hamiltonian is dominated by the kinetic term. It splits into a sum of terms, each one acting on one site alone. At each site, the two qubits can be in a singlet or triplet states; the lowest-lying state is the singlet and, therefore, the strong coupling ground state is simply a tensor product of spin singlets: $\ket{\Omega} = \bigotimes_{i=0}^{L-1} \ket{\psi^s}_{2i,2i+1}$ where $\ket{\psi^s} = \frac{1}{\sqrt{2}}(\ket{01}-\ket{10})$.
The first excited states are, in the strong coupling limit, obtained by replacing one of the singlets with a triplet state $\ket{\psi^t_m}$ with $m=1,0,-1$.
Even higher energy eigenstates can be found by replacing more singlets with triplets where the number of triplets determines the energies of the states.

The operators connecting a singlet to a triplet are: 
\begin{equation}
    O^\dagger_{x,m} = 
    \begin{cases}
        Z_{2x+1}, & m = 0, \\
        \pm\frac{1}{\sqrt{2}}(X\pm iY)_{2x+1}, & m = \pm 1, \\
    \end{cases}
\label{eq:one_pauli_ops}
\end{equation} where $x = 0,...,L-1$. 
These operators act exclusively on ``fuzz" qubits, though similar operators acting on only ``head" qubits work equally well.
As the Hamiltonian is $O(3)$ and translation invariant, states with different momenta and/or angular momenta do not mix.
Operators generating these states are
$O^\dagger_{k,m} = \frac{1}{\sqrt{L}}\sum_{x=0}^{L-1} e^{i k x} O^\dagger_{x,m}$, where $k = \frac{2\pi n}{L}$ with $n=0,1,\ldots,L-1$.
As we move away from the strong coupling limit, the potential term in \eq{eq:H} becomes relevant and  couples the ``head" qubits of nearest-neighbor sites. The resulting energy eigenstates are deformed from the strong coupling ones and the connecting operators are deformed away from \eq{eq:one_pauli_ops} as well. 

The opposite, weak-coupling limit of \eq{eq:H}, is also tractable. At $g^2\rightarrow 0$ the ``fuzz" decouples from the ``head" qubits, which themselves form a spin-$1/2$ anti-ferromagnetic Heisenberg chain. The Heisenberg chain can be solved by the Bethe ansatz method \cite{hulthen}. The ``fuzz" qubits are not coupled to the ``head" or to themselves and any of their states has zero energy. Consequently, the eigenstates in the weak coupling limit are the tensor product of eigenstates of the anti-ferromagnetic Heisenberg chain with any state of the ``fuzz" qubits:
\beq
\ket{\Omega}_{weak}
\approx
\ket{n}_{Heisenberg} \otimes \ket{any}.
\eeq These states can be used as approximations of the true ground state of \eq{eq:H} but the construction of $\ket{\Omega}_{weak} $ is expensive in terms of quantum gates and, given the limitations of available quantum computers, were not used in this work. However, we stress that it is common in field theories to have good approximations to the ground state in both the weak and strong coupling limits. 

\section{Numerical Results}
\label{sec:results}

\begin{table}
    \centering
    \begin{tabular}{|c|l|}
    \hline
        Set $1$ & $\{O^\dagger_{x,0}\}$, $\forall x$\\
        \hline
        Set $2$ & $\{O^\dagger_{x,0} , \frac{1}{\sqrt{2}}(O^\dagger_{x,1}O^\dagger_{y,-1}-O^\dagger_{x,-1}O^\dagger_{y,1})\}$ $\forall x<y$\\
        \hline
        Set $3$ & $\{O^\dagger_{x,m},O^\dagger_{x,m}O^\dagger_{y,m'}\}$ $\forall x<y,m, m'$ \\
        \hline
    \end{tabular}
    \caption{Operator sets used in the GEVP for the fuzzy $\sigma$-model in the strong coupling limit.}
    \label{tab:sets}
\end{table}

We explore now the method of \sect{sec:GEM} and probe how the unavoidable uncertainties arising from noise, Trotter errors, and finite shot number errors impact its effectiveness. We will use the fuzzy $\sigma$-model as our testing ground and perform exact calculations from the direct diagonalization of the hamiltonian, gate-based simulations and, runs on an actual quantum computer -- the IonQ Forte-Enterprise.

We start by performing exact calculations to extract the energy levels using three sets of operators summarized in \tab{tab:sets}. 
Set 1 contains the $m=0$ operators in \eq{eq:one_pauli_ops} at every site. 
Set 2 contains the product of 2 operators with quantum numbers $l=1,m=0$. 
Finally, set 3 contains all 1 and 2 site Pauli operators. The result of these calculations are shown in \fig{fig:gevp_combined}.

\subsection{Exact results for $L = 8$}

\begin{figure}[htbp]
    \centering

    \begin{minipage}[t]{0.48\textwidth}
        \centering
        \includegraphics[width=\linewidth]{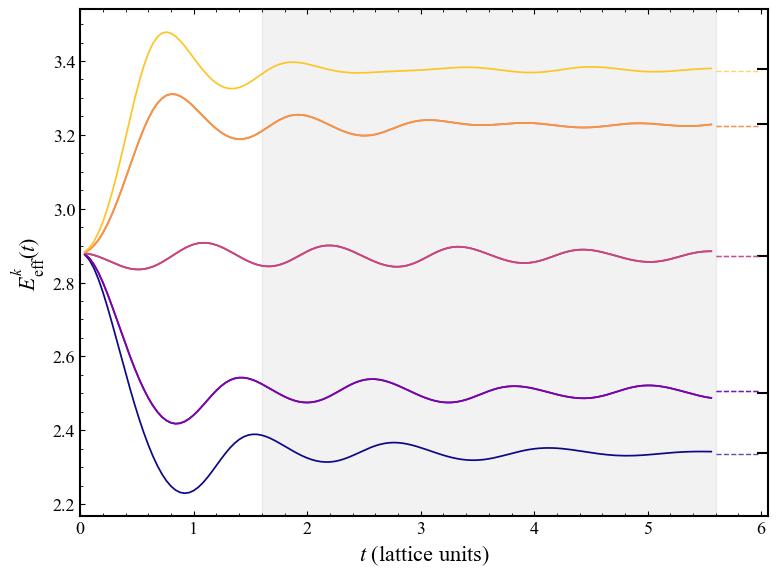}
    \end{minipage}\hfill
    \begin{minipage}[t]{0.48\textwidth}
        \centering
        \includegraphics[width=\linewidth]{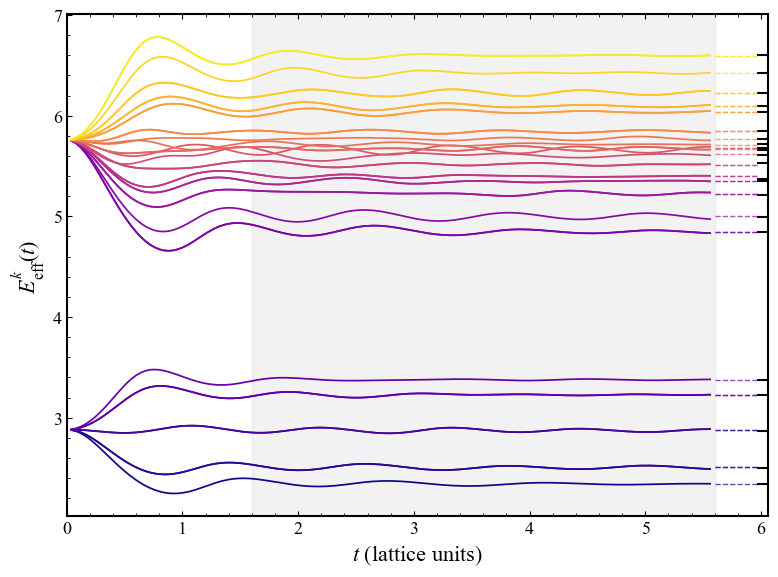 }
    \end{minipage}

    \vspace{0.8em}

    \begin{minipage}[t]{0.48\textwidth}
        \centering
        \includegraphics[width=\linewidth]{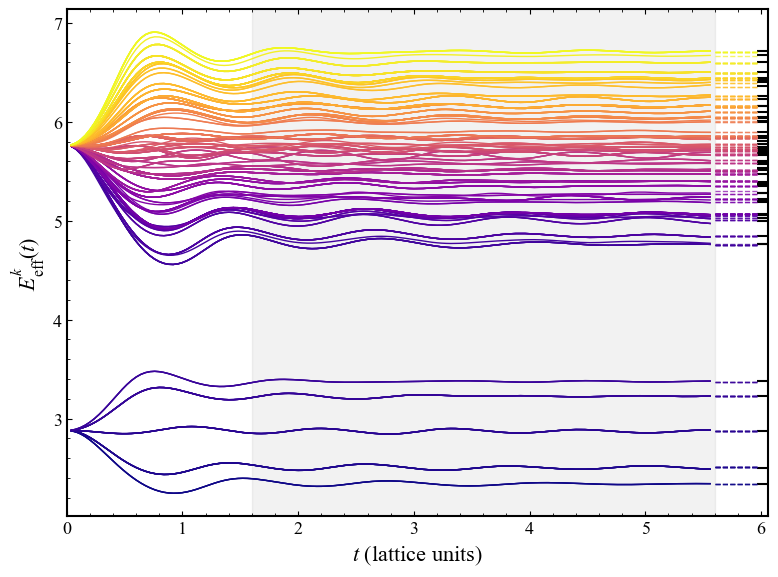}
    \end{minipage}\hfill
    \begin{minipage}[t]{0.48\textwidth}
        \centering
        \includegraphics[width=\linewidth]{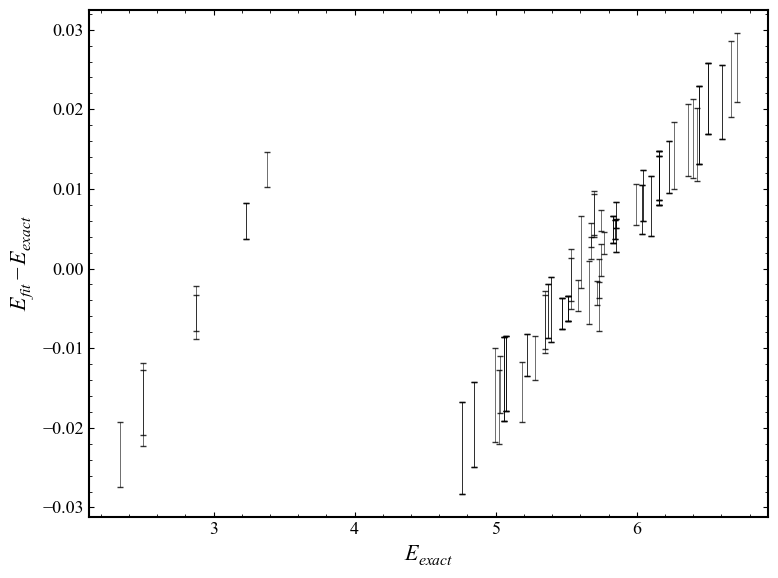}
    \end{minipage}

    \caption{
     Eigenvalues $\frac{i}{t}\log\lambda_i$ as a function of time for the $L=8$ fuzzy $\sigma$-model at $g=1.2$. 
        The grayed out area shows the region used for a fit to a constant.
        The dotted lines to the right of the gray region are the extracted energies 
        and the black, short lines on the right border are the exact energy levels. 
        (Top left) Eigenvalues obtained using operator set~1.
        (Top right) Eigenvalues obtained using operator set~2.
        (Bottom left) Eigenvalues obtained using operator set~3.
        The grayed-out region indicates the fit window used for a constant fit, while the dotted lines denote the extracted energy levels.
        (Bottom right) Residuals of the extracted energies, illustrating a systematic bias: low-lying states within a ``band" of states tend to be underestimated, while higher energies are overestimated.
    }
    \label{fig:gevp_combined}
\end{figure}

In order to separate the different sources of uncertainty, we apply our method to a noiseless system by constructing the exact time evolution operator $e^{-i H t}$ via directly diagonalization of $H$. We use a $L=8$ ($16$-qubit) system at $g=1.2$ and evaluate the correlators at 140 equally spaced steps with separation $dt=0.04$, corresponding to a maximum time evolution of $T_{\text{max}}=5.6$ in lattice units.
No Trotter or noise/decoherence  error is present and the only approximations are the finite
operator basis and the finite fit window ($t\in[1.6,5.6]$). 
The eigenvalues $\lambda_i$ of the correlator matrix \eq{eq:Cij} are then computed at each time $t$. 
According to \eq{eq:log-energies},  $\frac{i}{t}\log \lambda_i$  approaches the value of the energy levels at large $t$. 
The number of eigenvalues is given by the number of operators using in constructing $C_{ij}(t)$. A larger set of operators grants access to more  energy levels.   The results are shown in \fig{fig:gevp_combined}. The three sets of operators resolve $8$, $36$, and $276$ levels, respectively. In addition to the $\frac{i}{t}\log \lambda_i$ curves,
we show the energy values extracted by fitting the curves to a constant in the $t \geq 1.6$  region (dotted lines) and the exact energy levels (solid black ticks) obtained by exact diagonalization of the Hamiltonian. Even for the short times $t < 5.6$ considered here, we can extract the energy levels to sub-percent precision. For this particular Hamiltonian and the basis we used, the larger basis increased the precision of the middle range energy levels more than the lowest or highest lying levels. Notice also that the degeneracy pattern expected by the $O(3)$ symmetry of the model is reproduced by operator set 3. For instance, the two-Pauli operators in set 3 generate states with total spin $S=0,1,2$ as each Pauli operators has $l=1$, leading to one singlet, one triplet and one quintet, as observed in the data. 
This demonstrates that GEVP is able to pick up the proper combination of operators that are irreducible under $O(3)$. The errors shown at the bottom right of \fig{fig:gevp_combined} arise entirely from the finite time extent used in the fits used in the energy level extraction.

\begin{figure}
    \centering
    \includegraphics[width=0.5\linewidth]{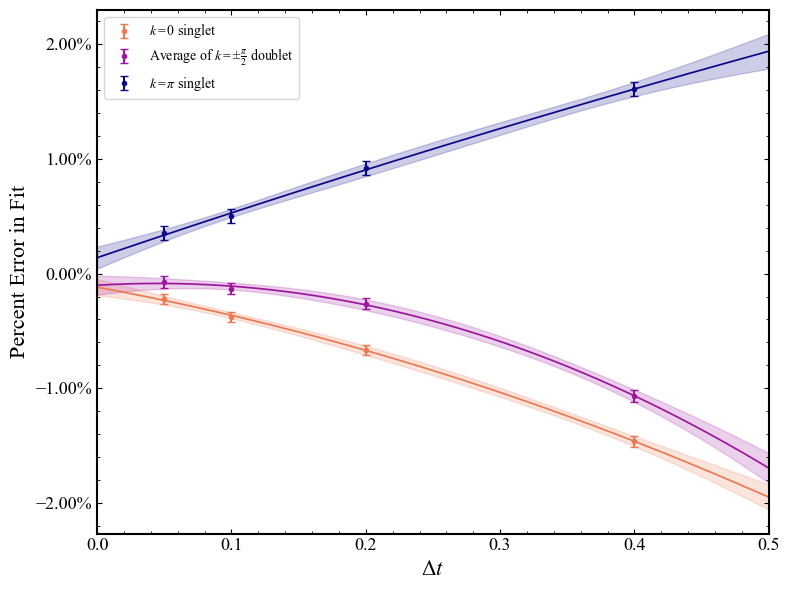}
    \caption {Relative difference between  extracted low-lying $l=0, m=0$ energy levels and exact ones. Here we use $L=4$, the operator set 1, and $T=5.6$. The bands show a quadratic fit and the bands indicate a one $\sigma$ statistical (not systematic) error. At any finite $\Delta t$ the lowest energy state ($k=\pi$) is systematically high while the highest energy state ($k=0$) is systematically low, leading to a narrowing of energy ranges. The two middle states ($k=\pm \pi/2$) are degenerate and so we only plot their average.
    The extrapolation to $\Delta t\rightarrow 0$ is consistent with the exact solution.
    }
    \label{fig:trotter}
\end{figure}

In a more realistic calculation, Trotter errors are unavoidable. In order to  gauge their impact, we perform a calculation on a $L=4$ lattice using a noiseless but gate-based calculation and a maximal time extent $T=5.6$. In \fig{fig:trotter} we show the energy level deviation of the exact result  on the Trotter step size $\Delta t$ and a quadratic extrapolation to $\Delta t\rightarrow 0$. We fit the energies (averaging over the doublet) at $\Delta t = 0.05,0.1, 0.2,0.4$ in to $f(\Delta t) = c_0+c_1\Delta t+c_2\Delta t^2$.

\begin{figure}
    \centering
    \includegraphics[width=0.5\linewidth]{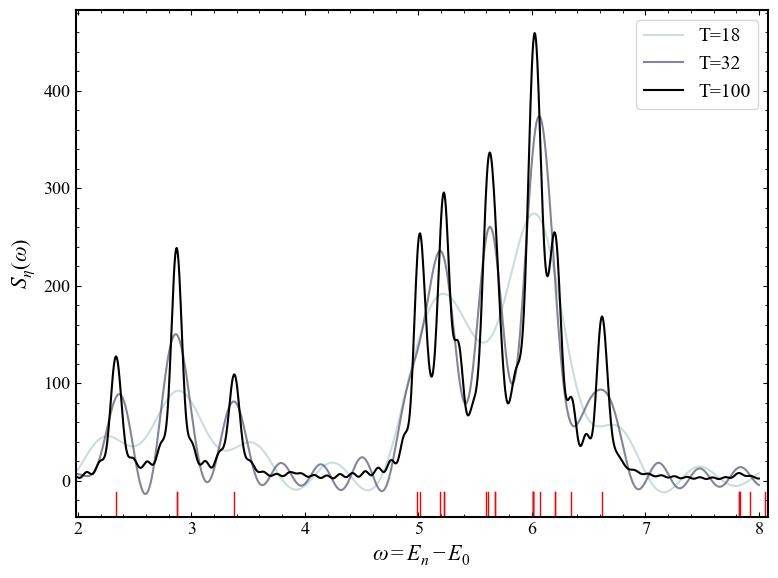}
    \caption{The convergence of the discrete Fourier transform method from \eq{eq:disc_ft} for $L=4$. We use an exponential damping factor of $\eta = 0.05$, as defined in \eq{eq:cont_ft}. We picked this value of $\eta$ to just barely smooth out the ringing artifacts of the $T=100$ curves. The red ticks indicate the exact energy levels. Even at $T=100$ several peaks are not resolved.}
    \label{fig:fourier}
\end{figure}

At this point, it is interesting to compare the efficiency of the GEVP method with a more obvious way of extracting energy levels, namely, Fourier transforming the correlators. 
Since the signal is a linear superposition of exponentials $e^{-i E_i t}$, the Fourier transform will exhibit peaks at frequencies $\omega \approx E_i$ that become sharper as the time extent $T$ increases. On \fig{fig:fourier} we show the result of this calculation with several values of $T$. As expected, the peaks are visible but require a much longer time extent $T$ to become well defined (notice that the smallest value of $T$ in \fig{fig:fourier} is over three times as large as the one in \fig{fig:gevp_combined} ). In a quantum calculation in a NISQ machine, where long time extents are impossible, this is a substantial advantage of the GEVP method.

\subsection{Quantum Computer $L = 4$ in quantum hardware}
\begin{figure}[t]
    \centering    \includegraphics[width=0.7\linewidth]{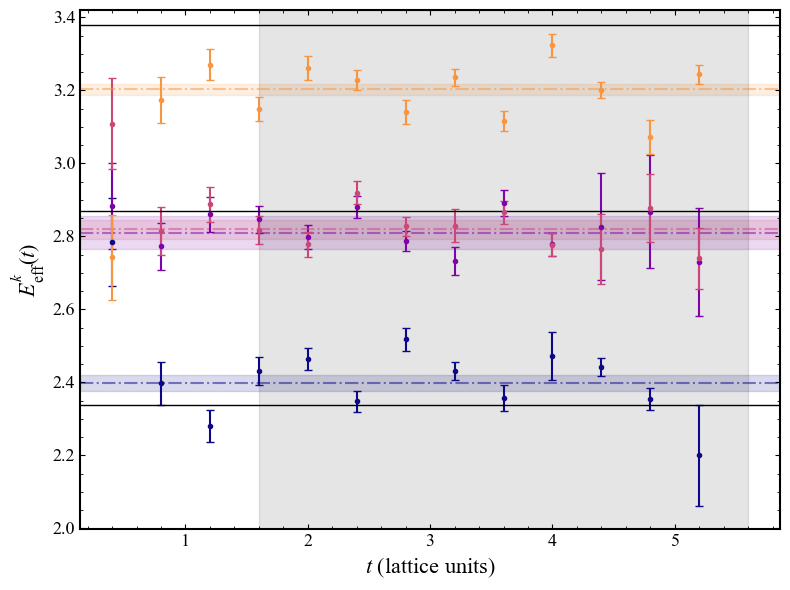}
    \caption{Result of run on IonQ's Forte Enterprise quantum computer using the smallest (set 1) of operators, $L=4$, $T=5.6$ and Trotter step $\Delta t=0.4$. Error bars reflect only the statistical errors from the finite number of shots (2048) and the bands show the  $1\ \Sigma$ uncertainty error in the fit to a constant. The exact energy levels are solid black lines. We can resolve the different energy levels and see the expected degeneracy.}
    \label{fig:qpu_L4}
\end{figure}

We have also tested our method on a quantum computer. We used the ion trap-based Forte-Enterprise from IonQ with a $L=4$ lattice, the same coupling $g=1.2$ as before, a trotterization step $\Delta t=0.4$, and a total of $14$ time steps.  Due to  cost, we restricted ourselves to the smaller operator set 1.
 The result is summarized in Figure~\ref{fig:qpu_L4} and Table~\ref{tab:qpu_L4_Z}. The quoted errors include only the statistical errors arising from the finite number of shots, estimated using the bootstrap procedure \cite{bootstrap}. The remaining small discrepancies are due to noise, Trotterization effects, and limited time extent.


\begin{table}
\centering
\begin{tabular}{|l|cccc|}
\hline
State & $E_{\mathrm{GEVP}}$ & $E_{\mathrm{exact}}$ & $\delta$ (\%) & $|\Delta|/\sigma$ \\
\hline
$k=0$ (singlet)      & $3.203(14)$ & $3.380$ & $5.3$ & $12.6$ \\
$k=\pm\pi/2$ (doublet)   & $2.810(46)$ & $2.870$ & $2.1$ & $1.3$ \\
  & $2.820(26)$ & $2.870$ & $1.7$ & $1.9$ \\
$k=\pi$ (singlet)        & $2.399(21)$ & $2.337$ & $2.6$ & $2.9$ \\
\hline
\end{tabular}
\caption{Hardware-extracted energies at $L=4$ using the $Z_{2x+1}$ basis
versus exact diagonalization. $\delta \equiv
|E_{\mathrm{GEVP}}-E_{\mathrm{exact}}|/E_{\mathrm{exact}}$;
$|\Delta|/\sigma$ is the deviation in units of the bootstrap error.
Parenthetical digits are bootstrap uncertainties on the final figures.}
\label{tab:qpu_L4_Z}
\end{table}

The $k=0$ singlet is the least accurately recovered level, 
displaced from the exact result by roughly thirteen times its statistical error. The $\Delta t \rightarrow 0$ extrapolation discussed in the previous section indicates that at $\Delta t=0.4$, Trotterization errors of the order of $2\%$ that increase the lowest energy and decrease the highest level are to be expected; we conclude that the remaining discrepancy is due to noise/decoherence.


\section{Conclusion}
\label{sec:conclusion}

We have shown that the generalized eigenvalue problem, standard practice for
Euclidean correlators in lattice field theory, adapts naturally to the real time
correlators produced by a quantum computer, and that it is a far more economical
way of extracting energy levels than Fourier transforming the same time series.
To demonstrate and test this method, we applied it to the fuzzy sigma model \cite{fuzzy}, a field theory that incorporates many features found in theories of interest.
By using the exact time evolution of the $L=8$ fuzzy $\sigma$-model we recovered energies
to sub-percent accuracy from a time extent of $T_{\rm max}=5.6$, with the
degeneracy pattern dictated by the $O(3)$ symmetry emerging automatically from the
diagonalization. On IonQ's Forte-Enterprise at $L=4$ the four levels accessible to
the single-Pauli basis were obtained within a few percent of their exact values,
with the residual discrepancies traced to a combination of Trotter and gate errors
rather than to the method itself.

The efficiency of the method rests on having a set of operators that approximately spans the
states of interest, and on an approximate ground state with good overlap with the
true one. Both were supplied here by the strong coupling limit of the model, but
neither is specific to it: any variational, dissipative, or adiabatic preparation of $|\Omega\rangle$
and any physically motivated operator basis can be used. This makes the procedure
a natural candidate for spectroscopy in theories where the sign problem obstructs
the Euclidean approach, in particular gauge theories at finite density and real
time transport.

\acknowledgments
P.B and V.S. were partially supported by 
U.S. DOE Grant
No. DE-SC0024286.

\bibliography{references}

\end{document}